# Phase transitions in multiferroic $BiFeO_3$ crystals, thin-layers, and ceramics: Enduring potential for a single phase, room-temperature magnetoelectric 'holy grail'

A. M. Kadomtseva[1], Yu.F. Popov[1], A.P. Pyatakov[1,2*] G.P. Vorob'ev[1], A.K. Zvezdin[2],

and D. Viehland[3]

[1] *M. V. Lomonosov Moscow State University, Leninskie gori, MSU, Physics department, Moscow 119992, Russia*
[2] *Institute of General Physics Russian Academy of Science, Vavilova st., 38, Moscow 119991, Russia*

[3] *Dept. of Materials Science and Engineering, Virginia Tech, Blacksburg, VA 24061*

Magnetic phase transitions in multiferroic bismuth ferrite (BiFeO3) induced by magnetic field, epitaxial strain, and composition modification are considered. These transitions from a spatially modulated spin spiral state to a homogenous antiferromagnetic one are accompanied by the release of latent magnetization and a linear magnetoelectric effect that makes BiFeO3-based materials efficient room-temperature single phase multiferroics.

Since the beginning of the multiferroic era (in the early 1960s), after Soviet scientists discovered a new class of materials that was called "ferroelectromagnets" [1], to our present time, bismuth ferrite $BiFeO_3$ has remained the prototypical example of a multiferroic: having a relatively simple structure and at the same time quite diversified and uncommon properties.

On the one hand due to its simple chemical and crystal structure, $BiFeO_3$ is a model system for fundamental and theoretical studies of multiferroics [2]. However, on the other hand, its unusual magnetic symmetry properties (space and time symmetry violation both in its crystal and magnetic structures) results in a variety of nontrivial consequences, including: (i) the unique coexistence of weak ferromagnetism and linear magnetoelectricity [3,4]; (ii) a toroidal moment, i.e., a special magnetic type of ordering [5-8]; (iii) the existence of an incommensurately modulated spin structure [9], previously only observed in magnetic metals; and (iv) magnetically induced optical second harmonic generation, observed for the first time in this particular material [10].

Furthermore, $BiFeO_3$ is the material with unique high ferroelectric Curie ($T_C$=1083 K) [11] and antiferromagnetic Neel ($T_N$=643K) [12] temperatures. However, its potential has yet to be realized, and might never be fully exploited. Difficulties persist as the magnetoelectric exchange and weak ferromagnetism are locked within a spin cycloid. A fundamental problem is that electronic configurations that favor magnetism are antagonistic to those that favor polarization [2] – compromise is necessary. Recently, investigations have shown that the multiferroic properties of $BiFeO_3$ can be dramatically increased by (i) epitaxial constraint [13], and/or (ii) rare earth substituents. These findings coupled with those in ME two-phase (nano and macro) composites of piezoelectric and magnetostrictive materials have served as triggers for a "magnetoelectric renaissance": the revival of hope to find a room temperature magnetoelectric material with significant coupling of polar and magnetic subsystems [14-16]. As a consequence, multiferroics are now being considered as promising materials for spintronics [17,18], magnetic memory systems, sensors, and tunable microwave devices [19]: offering the potential to revolutionize electromagnetic material's applications.

* Author to whom correspondence should be addressed: **alexander.pyatakov@gmail.com**

# Part I. BiFeO$_3$ single crystals: unrealized potential.

## I.1 Early Studies: Existence of both locally antiferromagnetic and long-range cycloid order

The crystal structure of bismuth ferrite is a rhombohedrally-distorted perovskite that belongs to the space group *R3c*. In Figure 1(a), the unit cell of BiFeO$_3$ is shown in its hexagonal representation ($[001]_{hex}$, $[100]_{hex}$, $[110]_{hex}$, $[010]_{hex}$ are hexagonal axis). Alternatively, in some cases, a pseudocubic representation has been used, where $[111]_c$ is equivalent to $[001]_{hex}$. Oxygen atoms (not shown) occupy face-centered sites of the Bi cubic framework.

BiFeO$_3$ has been shown to be ferroelectric with its polarization oriented along the rhombohedral c-axis (i.e., $[111]_c$) due to the displacement of Bi, and Fe, O relative to one another [20]; and shortly thereafter, neutron diffraction studies revealed antiferromagnetic (AFM) ordering along $[111]_c$ [21]. Spins in neighboring positions are antiparallel with each other, resulting in an AFM ordering of the *G*-type, as illustrated in Fig. 1(b).

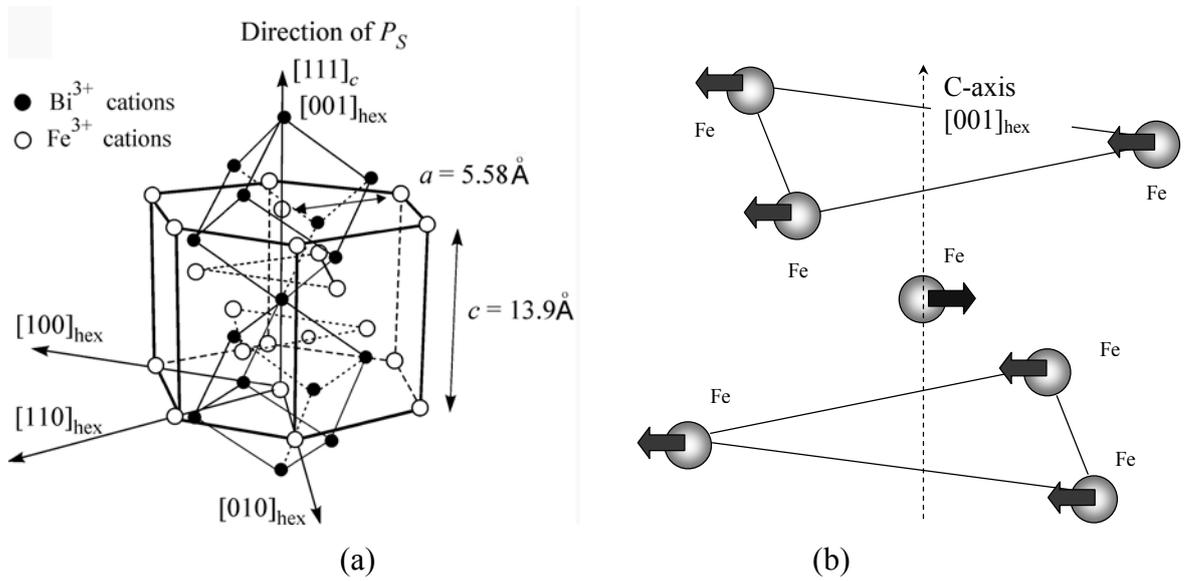

*Fig. 1 Crystallographic (a) and magnetic (b) structure of BiFeO$_3$*

In spite of a high Curie temperature [11] and a large polar displacement of ions [22], the measured value of the spontaneous polarization has been reported to be 6.1μC/cm$^2$ [11], which is surprisingly small compared to prototypical ferroelectrics such as PbTiO$_3$ (~100μC/cm$^2$). Interestingly, the magnetic symmetry of BiFeO$_3$ allows the linear magnetoelectric effect and weak ferromagnetism given in a thermodynamic potential as $E_i L_j H_k$ and $P_i L_j M_k$ terms. For the case of cubic symmetry these magnetoelectric terms are expressed in simple form:

$$F_{ME-induced} = -\alpha \cdot (\boldsymbol{E} \cdot [\boldsymbol{l} \times \boldsymbol{H}]); \qquad (1a)$$

$$F_{ME-spontaneous} = -\beta \cdot (\boldsymbol{P}_S \cdot [\boldsymbol{l} \times \boldsymbol{M}_S]); \qquad (1b)$$

where *α, β* are magnetoelectric constants, $\boldsymbol{P}_s$ is the spontaneous polarization, $\boldsymbol{l} = \boldsymbol{L}_l/2M_0$ the unit antiferromagnetic vector, $M_0$ the magnitude of the magnetization vector of the sublattices, $\boldsymbol{M}_s$ the in-plane spontaneous magnetization, and $\boldsymbol{E}$ and $\boldsymbol{H}$ the electric and magnetic fields. It is noteworthy that in contrast to conventional weak ferromagnetism that is expressed in Dzyaloshinskii-Moriya tems $L_i M_j$ [23,24], weak ferromagnetism is possible only in the presence of spontaneous polarization $P_s$: which is why $P_i L_j M_k$ terms (1b) are often referred to as *Dzyaloshinskii-Moriya-like*. Due to small in-plane anisotropy, the spontaneous magnetization in BiFeO$_3$ was expected to be easily controlled by a magnetic field of ~1*Oe*; furthermore, as weak ferromagnetism in BiFeO$_3$ is magnetoelectric (ME) in origin, a unique coexistence of weak ferromagnetism and linear magnetoelectricity – forbidden under conventional circumstances [3] – was expected. However, neither weak ferromagnetism nor linear magnetoelectric couplings were observed. Unfortunately, this fact has undermined the great potential of BiFeO$_3$ as an unusual ferroelectromagnetic material with high magnetoelectric coupling at high temperatures: which has been the lingering goal of scientists working in the field since the 1960s. Only since 2000 has single phase BiFeO$_3$ materials been developed with enhanced multiferroic properties in epitaxial thin-layer form [13] (see *Part II*), i.e., there persists promise.

The reason for the lack of weak-ferromagnetism and a linear magnetoelectric effect was found in the 1980's by precise time-of-flight neutron measurements [9]. This investigation revealed that the G-type AFM structure is not a complete description of the spin structure in BiFeO$_3$: rather, in addition, it has a long-wavelength modulation that forms a spin cycloid with a wave-vector that is oriented along [110]$_{hex}$, which is perpendicular to the [111]$_c$, as shown in Fig.2. This modulation results in a zero value for the volume-averaged ME effect and spontaneous magnetization. Thus, only a quadratic magnetoelectric effect whose value averaged over cycloid period is not zero was reported [25].

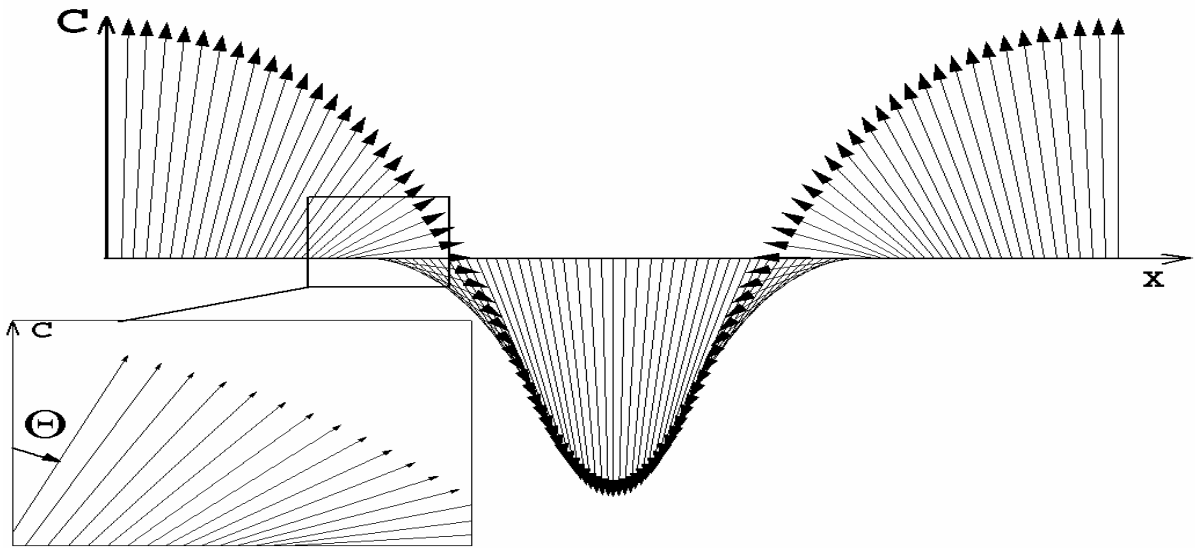

*Fig. 2 Long range magnetic order: incommensurate spin cycloid. The arrows correspond to the antiferromagnetic vector L that changes its orientation in space* $\theta = \theta(x)$.

The position of satellite peaks observed in time-of-flight neutron studies along the [110]$_{hex}$ (see Fig.3a) enabled determination of the period of the spin cycloid to be $\sim 620 \overset{\circ}{A}$. These results demonstrated that the periodicity of the cycloid was incommensurate with that of the lattice parameters. The presence of only a single diffraction peak along the (003) shows that there is no modulation along the [111]$_c$. It is noteworthy to mention that before these measurements spatially modulated structures were only observed in magnetic metals: so, the observation of a spin cycloid in magnetic dielectric BiFeO$_3$ was rather surprising. Additional experimental evidence of a spin cycloid was provided by nuclear magnetic measurements (NMR) in BiFeO$_3$ ceramic doped with $^{57}$Fe [26]. Instead of a single peak corresponding to a homogeneous structure, the NMR line was more complicated, exhibiting two maxima corresponding to the spin orientations perpendicular and parallel to [111]$_c$ (see Fig.3b). The NMR lineshape was asymmetrical at low temperature (*T*=4.2 K) and became increasingly symmetric at higher temperatures. This behavior was explained in terms of anharmonic cycloid: over most of the cycloid's period, spins are at a small angle (θ) with respect to [111]$_c$, as can be seen from the stronger intensity of the high frequency peak, and with increasing temperature the anharmonicity of the spin profile decreases resulting in nearly linear dependence of $\theta$ on coordinate along spin cycloid direction at room temperatures [26]. However in a recent report on a high resolution neutron powder diffraction study [27] it was stated that the character of the modulated cycloidal ordering of the $Fe^{3+}$ magnetic moments remains the same from 4 K up to the Neel temperature of 640 K. Thus the temperature behavior of the spin cycloid remains an intriguing question.

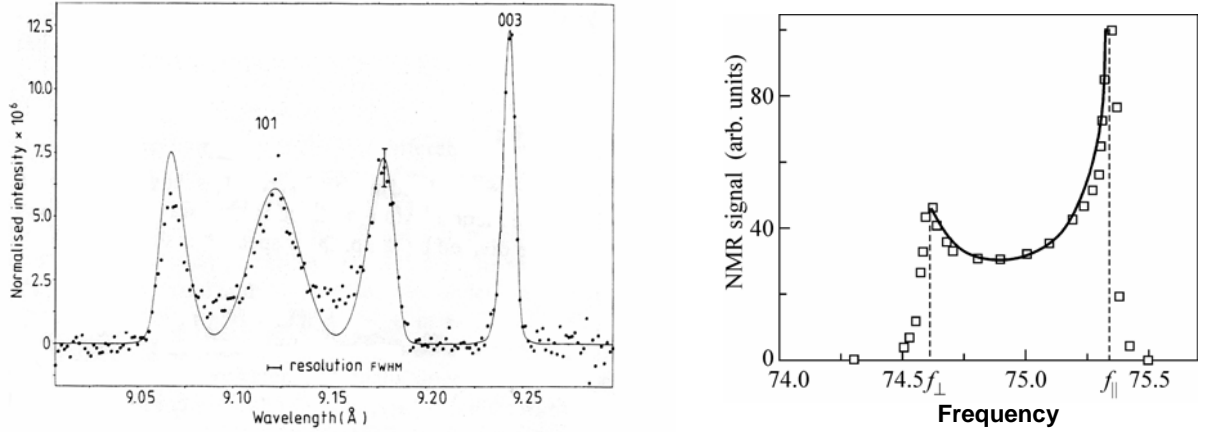

*Fig. 3 Experimental evidences for spin cycloid existence: (a) Time-of-flight neutron difraction measurements [9]; (b) Nuclear magnetic resonance measurements [26].*

The existence of a spin cycloid in BiFeO$_3$ has been explained in terms of a relativistic *Lifshitz-like invariant* $P_i L_j \nabla_k L_l$ [28-30]: named by analogy to the Lifshitz invariant $l_i \partial l_j / \partial x - l_j \partial l_i / \partial x$. For cubic symmetry, it takes the simple form of

$$F_{Lifshitz-like} = -\gamma \cdot P \cdot [L(\nabla L) - (L \cdot \nabla)L]; \qquad (2)$$

where γ is the coefficient of the inhomogenous magnetoelectric interaction. The omitted terms can be written as the total derivative $\nabla f(L)$, and do not contribute to the equation for the spin cycloid. Like the Dzyaloshinskii–Moriya-like exchange (1a), this inhomogenous interaction is magnetoelectric in origin and a spin cycloid is only possible in the presence of spontaneous polarization. It is remarkable that inhomogenous magnetoelectric interaction (2) can explain magnetoelectric effect in magnetic domain walls [31-33] and ferroelectricity in spiral magnets [34] under the assumption of polarization induced by magnetic inhomogeneity. Furthermore, there is a profound analogy between spatially modulated spin structures in multiferroics, and spatially modulated structures in nematic liquid crystals [35,36]. The periodic vector director structures in a nematic liquid crystal arise in an external electric field (i.e., a flexoelectric effect), and can be tuned under an applied electric field.

Thus, the inhomogeneous magnetoelectric interaction of equation (2) that gives rise to the cycloid prevents the observation of the homogeneous interaction of equation (1), which would manifest itself in a weak ferromagnetism (1a) and a linear magnetoelectric effect (1b). The necessary condition for observation of these two effects is the suppression of the spin modulated structure. Due to the magnetoelectric nature of the spin cycloid, it can be controlled by a magnetic and an electric field, and even destroyed under high field conditions that induce a homogeneous spin state (see *Section I.2*). Alternatively, the spatially modulated structure of BiFeO$_3$ can be disrupted by epitaxial constraint in thin-layers (see *Section II.1*) and/or by rare-earth substitutions (see *Section II.2*). Here, in the remaining parts of *Section I.1*, the theoretical framework for magnetoelectric interactions and spatially modulated spin structures in BiFeO$_3$ will be discussed.

*Symmetry and magnetoelectric interactions*

Magnetic symmetry is an elegant and efficient tool for understanding the physical properties of crystals with complex magnetic structures, notably antiferromagnets. On the basis of this approach, the Dzyaloshinski–Moriya interaction and weak ferromagnetism [23,24, 37,38], the linear magnetoelectric effect [39-41], piezomagnetism [42,43], linear magnetostriction [44], and a variety of unusual optical effects associated with the AFM vector (e.g., quadratic Faraday effect [45] or linear birefringence [46,47]) have been studied over many years.

The occurrence of one or the other of these effects in an AFM crystal with a centrosymmetric crystallographic structure depends on the space parity of its magnetic structure. For example, weak ferromagnetism arises in crystals with an even antiferromagnetic structure, whereas the linear magnetoelectric effect is forbidden in them. Conversely, a linear magnetoelectric effect is allowed and weak ferromagnetism forbidden in crystals with an odd antiferromagnetic structure. In this respect, crystals with trigonal symmetry – rhombohedral MnCo$_3$, FeBO$_3$, and α-Fe$_2$O$_3$ antiferromagnets with even magnetic structure, and the rhombhedral Cr$_2$O$_3$ antiferromagnet with an odd magnetic structure – have been studied most thoroughly. The crystal structures of all these materials possess an inversion center (space group $R\bar{3}c$). Here, in this subsection, the symmetry properties of BiFeO$_3$ are considered: it is akin to the aforementioned rhombohedral antiferromagnets, but yet different from them in that an inversion center is absent in both its crystal and magnetic structures. It will be shown that BiFeO$_3$ has unusual properties due to precisely this fact.

We choose the space group $R\bar{3}c$ as a "parent" symmetry for the $R3c$ symmetry under study. In reality, the phase transition in BiFeO$_3$ at the Curie point $T_C$ differs from a $R3c \rightarrow R\bar{3}c$ transition. However, this difference is not of importance in determining the adequate invariants that are responsible for the magnetoelectric properties of the system. The space group $R\bar{3}c$ differs from $R3c$ only by the presence of a polar vector **P**= (0, 0,$P_s$). Indeed, the parent symmetry can be used to develop the perturbation theory for determining the thermodynamic potential and other physical quantities of the system under the assumption of the smallness of **P**; i.e., we expand with respect to **P**. Correspondingly, $\xi = \Delta a / a$ is a small parameter, where $a$ is the lattice constant, and $\Delta a$ is the characteristic atomic deviation from the symmetric positions about the space inversion in $R3c$.

The magnetic exchange structure (i.e., the mutual directions of magnetic moments in the crystal) is determined by the following code using Turov's nomenclature [3]: $I^-$, $3_z^+$, $2_x^+$ where $I$ is the space-inversion element; $3_z$ (aligned with the $c$-axis) the threefold axis and $2_x$ the twofold axis are the group generators; and the indices ± of these elements that specify their parity about the transposition of magnetic sublattices (i.e., "+" indicates that the symmetry element transposes the ions within the same magnetic sublattice of an antiferromagnet, and "–" that the sublattice is transposed into one with opposite spin direction). Using these symmetry operators, the AFM vector can be shown to obey the following transformation rules:

$$I^\pm \boldsymbol{L} = \pm \boldsymbol{L},\ 2_x^\pm L_x = \pm L_x,\ 2_x^\pm L_{y(z)} = \mp L_{y(z)}.$$

For the other vectors, the action of elements with different indices is the same:

$$I^\pm \mathbf{m} = \mathbf{m},\ 2_x^\pm m_x = m_x,\ 2_x^\pm m_{y(z)} = -m_{y(z)},\ I^\pm \boldsymbol{P} = -\boldsymbol{P},\ 2_x^\pm P_x = P_x,\ 2_x^\pm P_{y(z)} = -P_{y(z)}.$$

The layout of the group elements relative to the magnetic ions in bismuth ferrite is shown in Fig.4. It is interesting to compare the code of BiFeO$_3$ with the codes of other antiferromagnetic compounds belonging to the same space group: for hematite ($\alpha - Fe_2O_3$) $I^+$, $3_x^+$, $2_x^-$, and for chromite (Cr$_2$O$_3$) $I^-$, $3_z^+$, $2_x^-$, whose exchange structures are illustrated in Figs.4b and 4c, respectively[1].

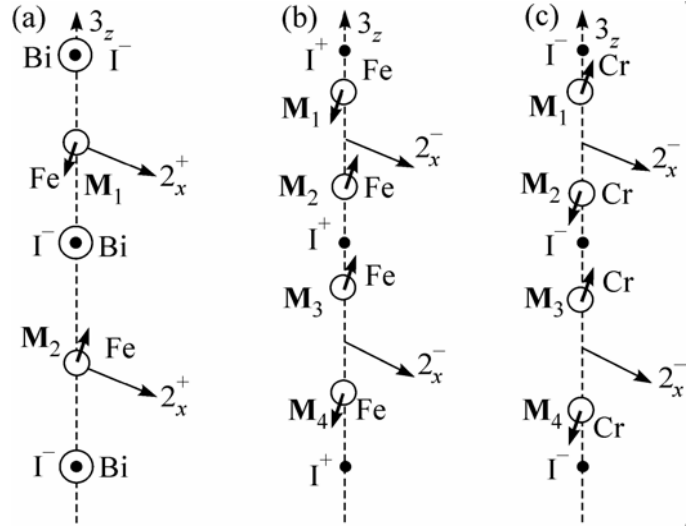

*Fig. 4. Exchange structures of (a) bismuth ferrite BiFeO$_3$, $L_1 = M_1 - M_2$; (b) hematite $L_2 = M_1 - M_2 - M_3 + M_4$; and (c) chromite Cr$_2$O$_3$, $L_3 = M_1 - M_2 + M_3 - M_4$.*

As crystal and magnetic cells are identical in BiFeO$_3$ the reduced group $\tilde{G}$ can be used where all translation on cell period are considered as unit operators. The space group contains eight irreducible representations: four one-dimensional ($\Gamma_1$, $\Gamma_2$, $\Gamma_4$, $\Gamma_5$) and two two-dimensional ($\Gamma_3$, $\Gamma_6$) ones (see Table I). Their matrix representations are given in the columns corresponding to the generating symmetry elements. The vector components of electric field $E$, magnetic field $H$, electric polarization $P$, magnetization $m$, and antiferromagnetic vectors $L_1$, $L_2$, and $L_3$ for the exchange structures of bismuth ferrite, hematite, and chromite, respectively (see Fig. 4), are given in the table according to their transformation properties. For instance, it follows from this table that the $(L_z)_1$ component changes sign and the vector $L_\perp = (L_x, L_y)_1$ transforms into $(L_x, -L_y)_1$ under the symmetry operation $2_x^+$. The transformation properties of the products $m_iL_i$, $H_iE_i$, $L_iE_i$ for bismuth ferrite are then given in Table II.

**Table I**

| | $E^+$ | $I^{+/-}$ | $3_z^+$ | $2_x^{+/-}$ | $E_i$ ; $P_i$ | $H_i$; $m_i$ | $L_i$ |
|---|---|---|---|---|---|---|---|
| $\Gamma_1$ | 1 | 1 | 1 | 1 | | | $(L_z)_2$ |
| $\Gamma_2$ | 1 | 1 | 1 | -1 | | $H_z$; $m_z$ | |
| $\Gamma_3$ | $\begin{pmatrix}1&0\\0&1\end{pmatrix}$ | $\begin{pmatrix}1&0\\0&1\end{pmatrix}$ | R | $\begin{pmatrix}1&0\\0&-1\end{pmatrix}$ | | $\begin{pmatrix}H_x\\H_y\end{pmatrix}$; $\begin{pmatrix}m_x\\m_y\end{pmatrix}$ | $\begin{pmatrix}L_y\\-L_x\end{pmatrix}_2$ |
| $\Gamma_4$ | 1 | -1 | 1 | 1 | | | $(L_z)_3$ |
| $\Gamma_5$ | 1 | -1 | 1 | -1 | $E_z$; $P_z$; $\nabla_z$ | | $(L_z)_1$ |
| $\Gamma_6$ | $\begin{pmatrix}1&0\\0&1\end{pmatrix}$ | $\begin{pmatrix}-1&0\\0&-1\end{pmatrix}$ | R | $\begin{pmatrix}1&0\\0&-1\end{pmatrix}$ | $\begin{pmatrix}E_x\\E_y\end{pmatrix}$; $\begin{pmatrix}P_x\\P_y\end{pmatrix}$; $\begin{pmatrix}\nabla_x\\\nabla_y\end{pmatrix}$ | | $\begin{pmatrix}L_x\\L_y\end{pmatrix}_1$; $\begin{pmatrix}L_y\\-L_x\end{pmatrix}_3$ |

*Table of irreducible representations of the reduced $R\bar{3}c$ space group (translation on cell period are considered as unit operators). Symbol R in table stands for operator of rotation at an angle of 120°*

---

[1] It is worth noticing that the exchange structure of BiFeO$_3$ still remains an open question. In the paper of Claude Ederer and Nicola A. Spaldin [Phys. Rev. B 71, 060401(R) (2005)] the position of inversion symmetry center is chosen at Fe atoms so the inversion element becomes $I^+$. The possible solution of this inconsistency can be found in the paper of R. de Suosa and Joel E. Moore arXiv:0806.2142 (this comment is made in 2008).

**Table II**

|  | $E^+$ | $\bar{I}$ | $3_z^+$ | $2_x^+$ | $m_i L_i$; $H_i L_i$; $H_i E_i$; $L_i E_i$; |
|---|---|---|---|---|---|
| $\Gamma_1$ | 1 | 1 | 1 | 1 | $(E_x L_x + E_y L_y)$ |
| $\Gamma_2$ | 1 | 1 | 1 | -1 | $(E_x L_y - E_y L_x)$ |
| $\Gamma_3$ | $\begin{pmatrix} 1 & 0 \\ 0 & 1 \end{pmatrix}$ | $\begin{pmatrix} 1 & 0 \\ 0 & 1 \end{pmatrix}$ | R | $\begin{pmatrix} 1 & 0 \\ 0 & -1 \end{pmatrix}$ |  |
| $\Gamma_4$ | 1 | -1 | 1 | 1 | $(m_x L_x + m_y L_y)$ |
| $\Gamma_5$ | 1 | -1 | 1 | -1 | $(m_y L_x - m_x L_y)$; $(H_y L_x - H_x L_y)$; $(H_y E_x - H_x E_y)$; |
| $\Gamma_6$ | $\begin{pmatrix} 1 & 0 \\ 0 & 1 \end{pmatrix}$ | $\begin{pmatrix} -1 & 0 \\ 0 & -1 \end{pmatrix}$ | R | $\begin{pmatrix} 1 & 0 \\ 0 & -1 \end{pmatrix}$ | $\begin{pmatrix} H_y L_y - H_x L_x \\ H_x L_y + H_y L_x \end{pmatrix}$ |

*Table of irreducible representations of the reduced $R\bar{3}c$ space group (translation on cell period are considered as unit operators). R in table stands for operator of rotation at an angle of 120°*

One can readily see in Table I that the $(H_x l_y - H_y l_x)$ combination in hematite, where **l** = $\mathbf{L}_2/2M_0$ is the unit antiferromagnetic vector and $M_0$ is the magnitude of the magnetization vector of the sublattices, corresponds to the first irreducible representation, i.e., its invariant. This invariant is responsible for the formation of a weak magnetization $(m_x, m_y) \sim (l_y, -l_x)$; i.e., the latter is a weak ferromagnet with the magnetization vector perpendicular to the antiferromagnetic one. It can also be shown from this table that a spontaneous magnetization is forbidden in $Cr_2O_3$. The combinations $m_j l_k$ (where $\mathbf{l} = \mathbf{L}_3/2M_0$) are not invariant. At the same time, the symmetry of $Cr_2O_3$ allows a linear magnetoelectric effect because of the $(E_x H_x + E_y H_y)l_z$, $E_z H_z l_z$, $H_z(E_x l_x + E_y l_y)$, and $E_z(H_x l_x + H_y l_y)$ invariants.

One can also easily find the following invariant for bismuth ferrite, using Tables I and II:

$$F = \ldots - P_z(m_y l_x - m_x l_y) + \ldots; \qquad (3)$$

where $P_z$ is the component of spontaneous polarization $\mathbf{P} = (0, 0, P_z)$ along the *c*-axis, $\mathbf{m} = (\mathbf{M}_1 + \mathbf{M}_2)/2M_0$ and $\mathbf{l} = \mathbf{L}_1/2M_0$ the unit magnetization and antiferromagnetic vectors, respectively, and $M_0$ the magnitude of the magnetization vector of the sublattices. This interaction is Dzyaloshinski–Moriya-like (1), and gives rise to a weak ferromagnetism with a magnetization of

$$\mathbf{m} \sim (P_z l_y, -P_z l_x, 0). \qquad (4)$$

This result is by no means contradictory to the well-known theorem in the theory of antiferromagnetism about the impossibility of weak ferromagnetism coexisting with a magnetoelectric effect [3]. Rather, this theorem is related to those antiferromagnets whose space group contains a space-inversion element (i.e., a crystal structure even about space inversion). For $BiFeO_3$, we deal with ferroelectromagnets where this requirement is not fulfilled. It is worth noting that the physical nature of weak ferromagnetism in a ferroelectromagnet is basically different from that of the Dzyaloshinski–Moriya case. A weak ferromagnetic moment in ferroelectromagnets results from magnetoelectric interaction: in other words, this magnetic moment arises due to the internal effective electric field. The expression for the free-energy density that includes the magnetoelectric terms proportional to the invariants of the $H_i E_i l_i$ type can be given as

$$f = ... - a_1 [E_x(H_y l_y - H_x l_x) + E_y(H_x l_y + H_y l_x)] \\ - a_2 H_z(E_x l_y - E_y l_x) - a_3 E_z(H_y l_x - H_x l_y) - a_4 l_z(H_y E_x - H_x E_y) + ... \quad (5)$$

The tensor relating the magnetic-field-induced polarization to the magnetic vector for the linear magnetoelectric effect is

$$\alpha_{ij} = \begin{vmatrix} -a_1 l_x & a_4 l_z + a_1 l_y & a_2 l_y \\ a_1 l_y - a_4 l_z & a_1 l_x & -a_2 l_x \\ -a_3 l_y & a_3 l_x & 0 \end{vmatrix}. \quad (6)$$

The coexistence of weak ferromagnetism (4) and magnetoelectric effect (5) not only remarkable itself but also may have an interesting consequence such as magnetoelectric effect, magnetic susceptibility, and electric polarizability enhancement as well as a renormalization of the values of the spontaneous electric polarization and magnetization [48].

Apart from the magnetoelectric (ME) effect and spontaneous magnetization, the magnetic symmetry of bismuth ferrite also allows for magnetic ordering of a special toroidal type [5,6]. The toroidal moment is a value conjugate to the product $[\mathbf{E} \times \mathbf{H}]$, and results from the $(\mathbf{T}[\mathbf{E} \times \mathbf{H}])$ term in the free energy expansion. It follows that the vector components of the toroidal moment are determined by the antisymmetric part of the linear ME tensor, given as:

$$T_i = \varepsilon_{ijk} \alpha_{jk}, \quad (7a)$$

$$T_1 = \alpha_{23} - \alpha_{32}, \; T_2 = \alpha_{31} - \alpha_{13}, \; T_3 = \alpha_{12} - \alpha_{21}; \quad (7b)$$

where $\varepsilon_{ijk}$ is the Kronecker tensor. By using (7b) and (6), one can readily verify that the vector components of the toroidal moment in BiFeO$_3$ are proportional to the components of the antiferromagnetic vector:

$$\begin{pmatrix} T_x \\ T_y \end{pmatrix} \sim \begin{pmatrix} l_x \\ l_y \end{pmatrix}; \; T_z \sim l_z. \quad (8)$$

The analysis of the irreducible representations of crystal space groups allows one to predict not only the tensor components of magnetic materials, but also their microscopic magnetic structure. Taking into account that the differential operator $\nabla$ transforms as a polar vector, we can include this new term in the expression for the free energy density (Table I), rewriting the free energy as:

$$F_L = \gamma \cdot P_z (l_x \nabla_x l_z + l_y \nabla_y l_z - l_z \nabla_x l_x - l_z \nabla_y l_y). \quad (9)$$

This form has a Lifshitz-like invariant (2), which sets up the thermodynamic conditions responsible for the spin cycloid.

Therefore, there are two types of ME interactions that must be considered in BiFeO$_3$: a homogeneous interaction characterized by the tensor of the linear ME effect (6), and an inhomogeneous interaction characterized by the $\gamma$ constant in (9). A distinctive feature of bismuth ferrite is the presence of a spontaneous polarization. Thus, a space-inversion element is absent in its crystallographic space group, and taken together with the space-parity violation in its magnetic structure results in the coexistence of weak ferromagnetism (4), magnetoelectric effect (6), a toroidal moment (8) and an inhomogeneous ME interaction (9).

*Spatially modulated spin structure*

The total expression for the free-energy density has the form

$$F = F_{exch} + F_L + F_{an}, \qquad (10)$$

where

$$F_{exch} = A \sum_{i=x,y,z}(\nabla l_i)^2 = A\left((\nabla \theta)^2 + \sin^2\theta(\nabla\varphi)^2\right) \qquad (11)$$

is the exchange energy, $A$ the constant of inhomogeneous exchange (or exchange stiffness), and $\theta$ and $\varphi$ are the polar and azimuthal angles of the unit antiferromagnetic vector $\mathbf{l} = (\sin\theta\cos\varphi, \sin\theta\sin\varphi, \cos\theta)$ in the spherical coordinate system with the polar axis aligned with the principal axis $c$, and where

$$F_{an} = -K_u \cos^2\theta \qquad (12)$$

is the anisotropy energy, and $K_u$ the anisotropy constant.

Minimization of the free-energy functional $F = \int f \cdot dV$ by the Lagrange–Euler method in the approximation ignoring anisotropy [30] gives for the functions $\theta(x,y,z)$ and $\varphi(x,y,z)$

$$\varphi_0 = const = arctg(\frac{q_y}{q_x}); \quad \theta_0 = q_x x + q_y y; \qquad (13)$$

where q is the wave vector of the cycloid. Equation (13) describes a cycloid whose plane is perpendicular to the basal plane and oriented along the propagation direction of the modulation wavevector.

The exact solution that takes into account anisotropy gives the following expressions for the spin distribution and cycloid period [49,50]:

$$\frac{d\theta}{dx} = \sqrt{\frac{K_u}{A\cdot m}}\sqrt{1-m\cos^2\theta} \qquad (14a)$$

$$\lambda = 4K_1(m)\sqrt{\frac{A\cdot m}{K_u}}; \qquad (14b)$$

where $K_1(m) = \int_0^{\pi/2}\frac{d\theta}{\sqrt{1-m\cos^2\theta}}$ is an elliptical integral of the first kind, and $m$ the modulus parameter of the elliptical integral that is found by minimization procedure of the free-energy (10) [50]. For an anisotropy constant much smaller than the exchange energy $K_u \ll Aq^2$, the modulus parameter $m$ tends to zero, and solution (14a) becomes harmonic with a linear dependence of $\theta$ on coordinates (13). By substituting (13) into (10), one can obtain the volume-averaged free-energy density for a harmonic cycloid approximation, as

$$\langle F \rangle = Aq^2 - (\gamma P_s)q - \frac{K_u}{2}. \qquad (15)$$

The wave-vector corresponding to the energy minimum is then

$$q_0 = \frac{2\pi}{\lambda} = \frac{\gamma \cdot P_s}{2A}. \tag{16}$$

Knowing that λ=620 Å [9], and assuming the polarization of BiFeO$_3$ to be $P_z = 6 \cdot 10^{-6} C/cm^2$ [11] and the exchange constant to be $A = 3 \cdot 10^{-7} \frac{erg}{cm}$ [12], one can estimate the inhomogeneous ME coefficient to be γ=10$^5$ erg/C=10$^{-2}$ V.

### I.2 High magnetic field studies. Field-induced transition: cycloidal →homogenous state

The magnetoelectric origin of the spin cycloid offers a means to control the magnetization state of BiFeO$_3$ by magnetic and/or electric fields. It is reasonable to suppose that an applied external magnetic and/or electric field above that of a critical value would result in the spin cycloid becoming energetically unfavorable. Above this critical point, a transition will occur to a homogenous antiferromagnetic spin state: consequently, all of the latent properties of bismuth ferrite hidden by the cycloid – such as linear magnetoelectric effect, spontaneous polarization, and toroidal moment – should become apparent. We will first show experimental evidence from a series of magnetic, magnetoelectric, magnetic resonance studies under high magnetic fields [4,7,28,29,51] that demonstrate such a spin transformation. Then, we will provide theoretical grounds for understanding the transition.

In Fig.5, we show the magnetization and polarization of BiFeO$_3$ induced under high magnetic fields. Measurements were performed at low temperature (T=10K) in order to suppress noise: as stated above similar magnetic and magnetoelectric properties should persist to room temperature. The experimental magnetization curves at low and high fields are well described by the linear dependences

$$\begin{cases} M = \chi \cdot H, \ H < 100 kOe & (17a) \\ M = M_{[001]}^{spont} + \chi_\perp H, H > H_c \ ; & (17b) \end{cases}$$

where $\chi = 5/6 \chi_\perp$; and $\chi_\perp = 5 \cdot 10^{-5}$ is the magnetic susceptibility in the direction perpendicular to the AFM vector *l*. However, abrupt changes can be seen in Fig.5 for both the magnetization and polarization near a critical field of $H_c$≈200kOe. We ascribe these changes to an induced transition between spatially-modulated and homogeneous spin states.

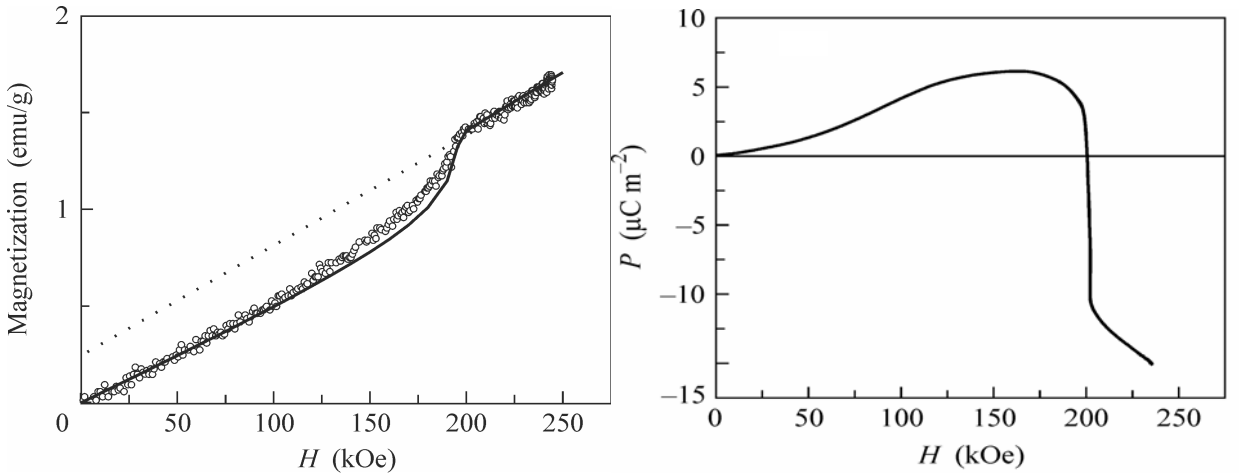

*Figure 5. (a) Magnetization of bismuth ferrite as a function of magnetic field [4]. Dots are the experimental data obtained in a field oriented along the [001]c direction, and solid line is the theoretical dependence (Eq. (29)); and (b) Dependence of the longitudinal electric polarization on the magnetic field for crystallographic direction along Bi atom cube edges [28].*

In Fig. 5a, weak ferromagnetism appears in the offset of the linear dependence of M on H in the field range of H>$H_c$ (17b). The value of spontaneous magnetization in this weak ferromagnetic state can be found by extrapolation of the high-field linear slope of the M-H curve to the ordinate axis: $M_{[001]}^{spont} \approx 0.25 emu/g$ (see Fig.5a). Taking into account the orientation of the crystal (the magnetization was measured along [001]$_c$). We can arrive at a value for the in-plane spontaneous magnetization of $M_s = \sqrt{\frac{3}{2}} 0.25 emu/g = 0.6 emu/g \approx 5G$ [4]. The changes in the magnetization curve in the vicinity of $H_c$ reflect the cycloidal→homogeneous transformation, which we will treat in more theoretical detail below in this section.

In Fig. 5b, a linear magnetoelectric effect can be seen from the linear dependence of P on H in the field range of H>$H_c$. For H<$H_C$ the electric polarization has a near-quadratic dependence on H; whereupon, near $H_c$, it undergoes an abrupt jump that is accompanied by the onset of a linear dependence on H. The value of the linear ME effect can be determined from the slope of the P-H curve at H>$H_c$, and can be found as ~$10^{-10}$ C/(m$^2$Oe) [28].

To determine the toroidal moment in BiFeO$_3$, the antisymmetric part of the ME tensor (6) was examined at fields exceeding $H_C$ for orientations at an angle of 45° to the a- and b-axes of the basal plane. Using such a field orientation, it is possible to measure both polarization components $P_a$(H) and $P_b$(H). A nonzero antisymmetric component in the magnetoelectric tensor (6) appears for H>$H_c$, evidencing a toroidal moment $T_z \sim (\alpha_{12} - \alpha_{21})$ upon inducing the homogeneous spin state [4,7].

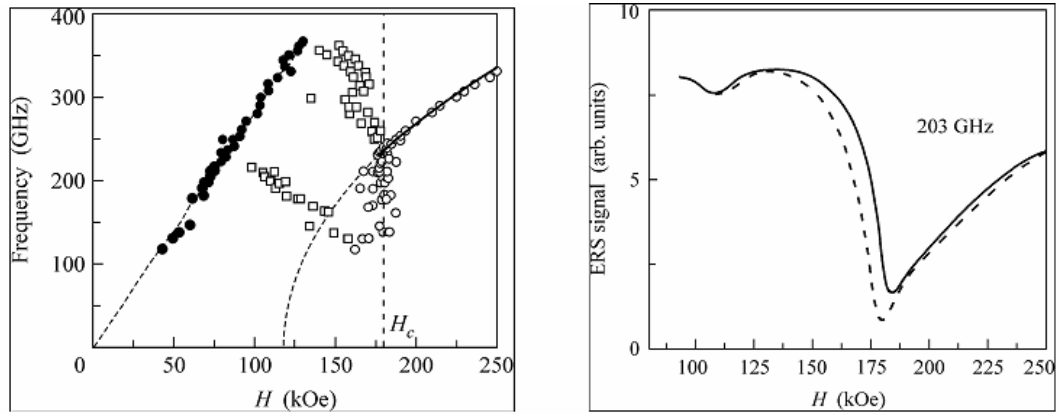

*Figure 6 (a) Antiferromagnetic resonance frequencies as functions of magnetic field H [51]; (b) Magnetic hysteresis of the absorption peak. Solid line is for the increasing field and dashed one is for the decreasing field [51].*

All of the abovementioned high-field measurements [4,7,28,29] of macroscopic property changes associated with the induced transition were performed using pulsed fields. However, local probes have also been used to investigate changes in electron spin resonance (ESR) spectra of BiFeO$_3$ [51] under static magnetic fields. Measurements were done at liquid helium temperatures for the same reasons as those made under pulsed field (i.e., reduce noise) at the National High Magetic Field Laboratory (Tallahassee, FL). Figure 6a shows the frequency of the ESR signal as a function of H for 50<H<250kOe. Strong changes in the resonance frequency were observed near a critical field of $H_c$ =180kOe. A Landau-Ginzburg phenomenological theory was developed to theoretically model the electron spin resonance spectra in the homogenous spin state (H>$H_c$) by taking into account the Dzyaloshinski-Moriya-like magnetoelectric interaction (4), represented in Fig.6a by a dashed line. Fitting of the data to the phenomenological model yielded a value of $M_s$=6G for the spontaneous magnetization, in agreement with the direct measurements of the magnetization changes under a pulsed field [4]. Also, please note that the induced phase transformation was accompanied by appreciable hysteresis in the resonance spectra between increasing and decreasing field strengths, as shown in Fig. 6b: evidencing a 1st order type transition.

The combinations of the data in Figures 5 and 6 demonstrate agreement between macroscopic property changes and those made using a local probe: both revealing an induced phase transition near a critical magnetic field. The results are in agreement with respect to the critical field levels required to induce the transformation, and induced magnetization and polarization changes. In what follows, we will discuss the thermodynamic framework which explains these observations based on an induced transformation between cyloidal and homogeneous spin states.

**Theory of magnetic and electric field-induced phase transitions**

Application of high magnetic and electric fields will result in changes in the effective anisotropy of BiFeO$_3$, i.e., $K_{eff}(E,H)$. This tunability of the uniaxial anisotropy constant K$_{eff}$ by E or H offers the possibility of a spin cycloid transformation, in accordance with (14). Consider the case of ***H***||***E***||*c*. It is convenient to use dimensionless units of electric and magnetic fields:

$$e = \kappa_\| \frac{E_z}{P_s}, \tag{18 a}$$

$$h = H_z \sqrt{\frac{\chi_\perp}{2Aq_0^2}}; \tag{18 b}$$

where $\kappa_\|$ is the electric susceptibility in the direction perpendicular to the c-axis, $\chi_\perp$ the magnetic susceptibility in the direction perpendicular to the AFM vector ***l***, $q_0$ (16) the value of the wavevector of the cycloid corresponding to the minimum of the free energy (10) in the absence of external fields and neglecting anisotropy. The electric polarization of the medium and the magnetization of the magnetoelectric origin in the linear approximation can be represented in the form:

$$P_z = P_s(1+e) \tag{19 a}$$

$$M = M_s(1+e); \tag{19 b}$$

where $M_s = \chi_\perp DP_s$ is the spontaneous magnetization due to a Dzyaloshinskii-Moriya-like interaction (1a), and $D$ is a homogenous magnetoelectric interaction constant.

The free energy density can be conveniently written as the sum

$$f = f_{exch} + f_L + f_{an}, \tag{20}$$

where the energies of exchange, inhomogeneous magnetoelectric interaction, and effective anisotropy are normalized to the exchange energy $Aq_0^2$ of the harmonic cycloid in the absence of applied fields: that is,

$$f_{exch} = \frac{F_{exch}}{Aq_0^2} = \frac{1}{q_0^2}\left(\frac{d\theta}{dx}\right)^2 \tag{21}$$

$$f_L = \frac{F_L}{Aq_0^2} = -\frac{2|1+e|}{q_0} \cdot \left|\frac{d\theta}{dx}\right| \tag{22}$$

$$f_{an} = -k(e,h)\cos^2\theta; \tag{23}$$

where $k(e,h) = \dfrac{K_{eff}}{Aq_0^2} = \left(k_u - h^2 - \beta\cdot(1+e)^2\right)$ is the dimensionless effective anisotropy constant, which accounts for the effects of electric $e$ and magnetic $h$ fields. For the homogeneous state with an AFM unit vector $l \perp c$ ($\theta = \dfrac{\pi}{2}$) we have

$$f_\perp = 0. \tag{24}$$

As for the homogeneous state $l \parallel c$ ($\theta=0$), it was shown in [50] that the parallel phase is only possible under electric fields applied antiparallel to $P_s$, which are much larger than the ferroelectric coercive field of $BiFeO_3$; and thus unlikely to exist for this particular material.

In the spin cycloid state, we can obtain the expression for the total free energy (20) averaged over the period $\langle f \rangle_\lambda = \dfrac{4}{\lambda}\int_0^{\pi/2} f(\theta)\dfrac{dx}{d\theta}d\theta$ by taking into account (14a):

$$\langle f \rangle_\lambda = -\dfrac{k(e,h)}{m}\left(1 - 2\dfrac{K_2(m)}{K_1(m)}\right) - \dfrac{\pi}{K_1}\sqrt{\dfrac{k(e,h)}{m}}|1+e|; \tag{25}$$

where $K_1(m) = \int_0^{\pi/2}\dfrac{d\theta}{\sqrt{1-m\cos^2\theta}}$, $K_2(m) = \int_0^{\pi/2}\sqrt{1-m\cos^2\theta}\cdot d\theta$ are elliptical integrals of the first and second kind, respectively. To each pair of field strengths ($e, h$) there corresponds a modulus $m$ of the elliptic integral for which the energy is minimum. Physically, this means that the cycloid's profile changes under applied fields. Under strong fields, its shape differs significantly from that of a harmonic profile, becoming similar to a function describing a periodic structure of domains separated by walls (solitons) whose widths are considerably smaller than the domain width. It follows from (24) that, upon the transformation to the phase with the AFM unit vector $l \perp c$, the energy of the domain walls $\langle f \rangle_\lambda$ changes sign and the spatially modulated spin state becomes energetically unfavorable. The zero-order approximation then corresponds to a harmonic cycloid with a linear spin distribution (13), which upon taking into account (20-23), has the simplified form

$$\langle f \rangle_\lambda = -(1+e)^2 - \dfrac{k(e,h)}{2}. \tag{26}$$

Then, by using (24) and (26), we can obtain analytical expressions corresponding to the phase boundary between the spatially modulated and the homogeneous states as

$$h^2 = k_u + 2(1+e_z)^2\left(1 - \dfrac{\beta}{2}\right). \tag{27}$$

Using (27), the phase diagram can be constructed according to the results of numerical calculations, as shown in Figure 7. The energy of the spatially modulated structure (25) was minimized with respect to the modulus $m$ at each point ($e, h$). The calculated energy was compared with the energy of the homogeneous state (24). In the calculations, the anisotropy parameter, the magnetoelectric interaction parameter, and the coercive force were taken as $k_u = 2$, $\beta = 1,2$, and $e_c = 0,04$, respectively. The material constants for $BiFeO_3$ corresponding to these values are: $K_u = 6\cdot 10^5$ erg/cm$^3$, $A = 3\cdot 10^{-7}$ erg/cm, $q_0 = 10^6$ cm$^{-1}$, $\chi_\perp = 5\cdot 10^{-5}$, $M_s = 6$ Gauss, $P_s = 0,06$ C/m$^2$, $E_c = 50$ kV/cm, and $\kappa_\parallel = \varepsilon - 1 \approx 50$ [50].

In Figure 7, the regions *I* and *II* are separated by a boundary shown as a solid line. This boundary corresponds to that between the homogeneous (with $l \perp c$) and the spatially modulated spin states. The limit of the spatially-modulated spin phase in the harmonic approximation according to expression (27) is indicated by a boundary represented as a dashed line. It can be seen that, owing to the change in the shape of the cycloid (characterized by the modulus *m* of the elliptic integrals), a solitonic cycloid can exist in the regions in which the harmonic cycloid is energetically unfavorable (the region between dashed and solid lines). The critical magnetic fields $h_c$ for numerical and analytical (27) solutions differ by ~10% at zero electric field; and, this discrepancy increases slightly with increasing electric field. In the absence of an applied electric field, the phase transition occurs at a point h=1.8, which corresponds to a magnetic field of ~190kOe. This is in good agreement with the experimental data reported in [4,7,28,29,51]. At the critical magnetic field, the phase transition can be shifted by h=0.1 (10kOe) upon simultaneous application of an electric field of 0.08 (100kV/cm). Figure 7 also illustrates the spin cycloid transformation under dually-applied electric and magnetic fields: the period of the cycloid decreases with increasing electric field; and at the field region between dashed and solid lines, the cycloid is domain-like and described by a soliton (14a) with $m \rightarrow -\infty$ at $h \rightarrow h_C$.

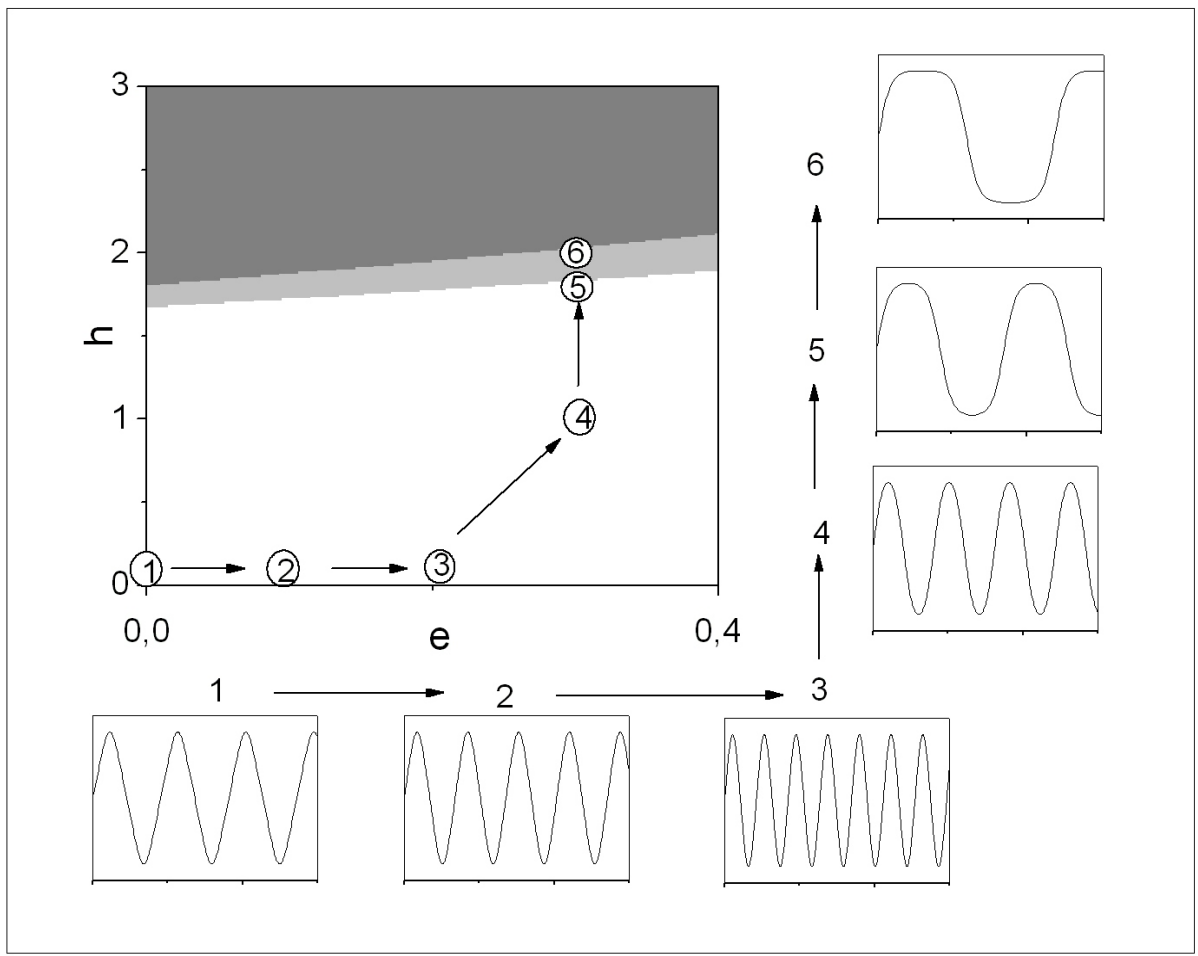

*Fig 7 Electric field-magnetic field phase diagram: e dimensionless electric field: 0.1 correspond to =125kV/cm; h dimensionless magnetic field: h=1 correspond to 110kOe; (1-6) fragment x-projection of antiferromagnetic vector on coordinate along direction of cycloid propagation (1-4) quasiharminic cycloid, (5,6) anharmonic (domain-like) cycloid.*

The anharmonic character of the cycloid in high magnetic fields becomes apparent in the nonlinear behaviour of the magnetization curve (see Fig.5a) for 150<H<200kOe [4]. This observed experimental dependence (17) can be understood in terms of the sum of spontaneous magnetization $M^{spont}$ averaged over the cycloid's period, and the magnetization $M^H$ induced by external field dependence on external field **H**:

$$\boldsymbol{M}(H) = \langle \boldsymbol{M}^{spont} \rangle(H) + \boldsymbol{M}^H(H). \qquad (28)$$

Experimental configurations somewhat more complex than that in the case of the phase diagram of Fig.7 can be treated: where the field is aligned with the $[001]_c$, or $\mathbf{H} = \left(0, H\sqrt{2}/\sqrt{3}, H/\sqrt{3}\right)$ in hexagonal representation. The total magnetization along $[001]_c$ in this case is

$$M_{[001]_c}(H) = M_{[001]_c}^{spont}\langle\sin\theta(x)\rangle_\lambda + \underbrace{\frac{2}{3}\chi_\perp H}_{M^H_{y[001]_c}} + \underbrace{\frac{1}{3}\chi_\perp H\langle\sin^2\theta(x)\rangle_\lambda}_{M^H_{z[001]_c}}; \qquad (29)$$

where θ is the angle between the spontaneous polarization $P_z$ (c axis) and the antiferromagnetic vector; and $M^H_{y[001]_c}$ and $M^H_{z[001]_c}$ are the projections of the y- and z- components of the induced magnetization $M^H$ onto the $[001]_c$, respectively. Equation (29) can then be used to explain the observed dependences of the spin profile: in the field range of $H \in (0, H_c)$, the spin profile changes from quasi-harmonic to that of a single domain antiferromagnetic state, as shown in Figure 8. Thus, averaged over an entire period $\frac{1}{\lambda}\int_0^{2\pi} f(x)\frac{dx}{d\theta}d\theta$, the values of sine $\langle\sin\theta(y)\rangle_\lambda$ and the square of sine $\langle\sin^2\theta(y)\rangle_\lambda$ vary within intervals of (0, 1) and (1/2, 1), respectively (fig.8). In turn, the dependence of (29) turns to that of (17a) at $H \to 0$, and to that of (17b) at $H \to H_c$.

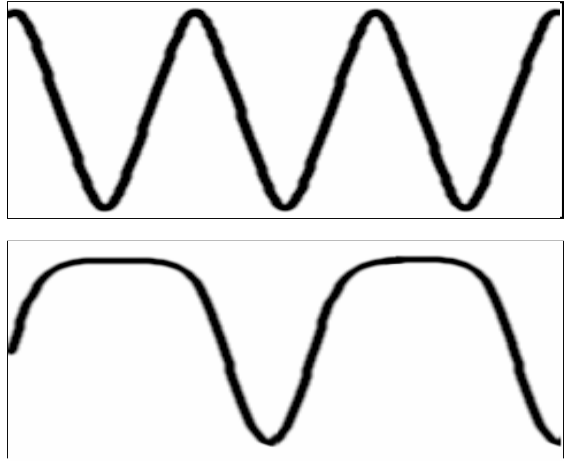

**Low field**
$\langle\sin\theta(y)\rangle_\lambda = 0$
$\langle\sin^2\theta(y)\rangle_\lambda = \frac{1}{2}$

**High field**
$\langle\sin\theta(y)\rangle_\lambda \to 1$
$\langle\sin^2\theta(y)\rangle_\lambda \to 1$

*Fig 8. The transformation of the cycloid in the high magnetic field directed along $[001]_c$ axis (i.e. magnetic field has nonzero projection on the basal $(001)_{hex}$ plane).*

*Summary*

In spite of many investigations spanning nearly 50 years, multiferroic properties in $BiFeO_3$ crystals have not been realized: except under extremely high fields. In a nutshell, this disappointment is due to the trapping of the magnetic vector and linear magnetoelectric interactions within a spin cycloid: which can be destabilized at high fields. It is fundamental that electronic configurations that favor magnetism are antagonistic to those that favor polarization [2] – compromise is necessary. But, how?

# Part II. Epitaxial films, and rare-earth substitutents: Lingering potential

As we learned above in *Part 1*, the necessary condition to manifest multiferroic properties in bismuth ferrite is the suppression of the spin cycloid. Apart from application of an extremely high magnetic field of ~200kOe, other means exist of suppressing the spatially modulated structure in BiFeO$_3$. Recently, the mentioned suppression of the spin cycloid – accompanying by weak ferromagnetism and linear magnetoelectricity – has been observed in bismuth ferrite epitaxial thin-layers, and by rare-earth substituted crystalline solutions; as we will discuss in more detail below. These approaches offer hope that the special properties of BiFeO$_3$ may still be realized.

## II.1 Epitaxial thin-layers: Release of latent magnetization and observation of large polarization.

Epitaxial thin-films of BiFeO$_3$ have been grown on (001)$_c$ SrTiO$_3$ [13]. The P$_s$ of (001)$_c$ BiFeO$_3$ thin films (50-500nm) is ~60 C/m$^2$ – which is ~20x larger than that of a bulk crystal projected onto the same orientation. Hetero-epitaxy induces significant and important structural changes [52-54]. Synchrotron studies [54] on BiFeO$_3$ thin films of have shown that the monoclinic M$_a$ phase is stable for (001)$_c$ epitaxial films, with lattice parameters of $(a_m/\sqrt{2}, b_m/\sqrt{2}, c_m; \beta)$ = (3.907, 3.973, 3.997Å; 89.2°), which are distinctly different from the rhombohedral ones of bulk single crystals and ceramics of (a$_r$=3.96 Å). However, the studies have also shown that the rhombohedral phase is stable for (111)$_c$ epitaxial films [52], whose lattice parameters are identical to those of the bulk crystal. Clearly, the structure of BFO films can be engineered by epitaxial constraint.

Correspondingly, the crystallographic orientation of the film state has significant effect on the physical properties of BiFeO$_3$. Remanent polarizations P$_r$ of ~55 µC/cm$^2$ for (001)$_c$ films, ~80 µC/cm$^2$ for (110)$_c$ films, and ~100 µC/cm$^2$ for (111)$_c$ films have been reported [52]. Pulsed polarization studies have confirmed that this large value is due to a true dielectric displacement, without notable contributions from conduction. Fig. 9a shows $\sqrt{3}\cdot$P$_{(001)c}$, $\sqrt{2}\cdot$P$_{(110)c}$, and P$_{(111)c}$ as a function of E for the variously oriented films. In this figure, the values of the projected polarizations can be seen to be nearly equivalent. This reveals that the direction of spontaneous polarization lies close to (111)$_c$, and that the values measured along (110)$_c$ and (001)$_c$ are simply projections onto these orientations. Clearly, similar to bulk crystals and ceramics, the spontaneous polarization is oriented close to (111)$_c$, but yet is dramatically increased relative to the reported bulk values. The question still remains whether this enhancement for epitaxial thin layers is due to the constraint (as was theoretically justified in [55]), or rather simply due to the lower leakage currents and higher dielectric breakdown strengths which can allow for the observation of polarization switching in samples with large coercive fields. The studies of electrical and conductive properties of bismuth ferrite films [56-58] have shown that oxygen vacancies are the main cause for the high leakage current in BiFeO$_3$, while doping of higher valence ions such as Ti$^{4+}$ [57] or substitution Bi with Tb ions [58] reduces the dc conductivity by more than three order of magnitude down to $10^{-11}$ Ω$^{-1}$ cm$^{-1}$.

The M-H properties of the variously oriented films have also recently been characterized [53], using a superconducting quantum interference device (SQUID). The magnetization of the films at low magnetic field levels has been found to be dramatically increased, relative to that of bulk crystals. This is illustrated in Fig. 9b for a (111)$_c$ film, where it can be seen that the induced magnetization is on an order of 0.6 emu/g at H$_{ac}$=1Tesla – more than one order of magnitude higher than that of bulk crystals under the same field.

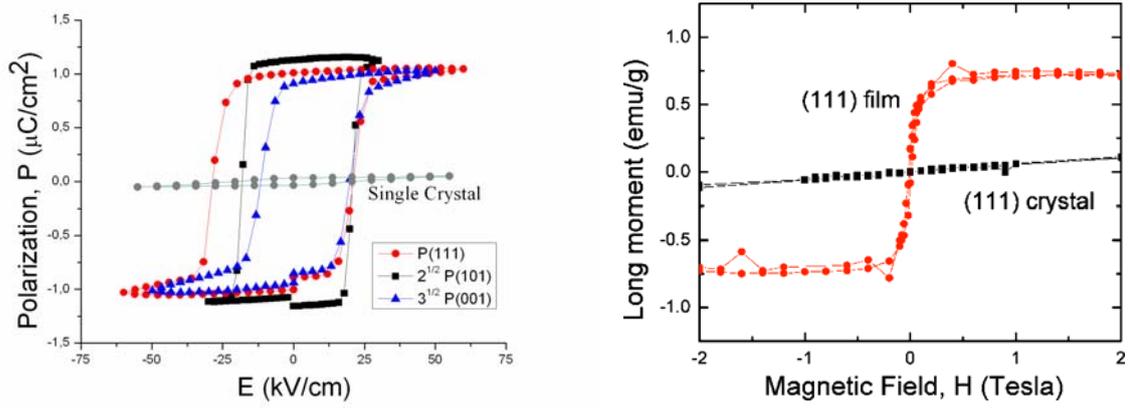

*Figure 9. (a) P-E response for BiFeO$_3$ epitaxial thin layers grown on (001)$_c$, (110)$_c$, and (111)$_c$ SrTiO$_3$ substrates by pulsed laser deposition; and (b) M-H response for a BiFeO$_3$ epitaxial thin layer grown on (111)$_c$ SrTiO$_3$, plotted next to that of a (111)$_c$ oriented BiFeO$_3$ bulk single crystal [53].*

It should be noted that (i) the value of the saturation magnetization for the (111)$_c$ film is close to the spontaneous magnetization observed in bulk crystals at the cycloidal to homogenous antiferromagnetic spin transition field of H≅18 Tesla [4, 51]; and (ii) the structure of (111)$_c$ BiFeO$_3$ thin films is identical to that of bulk crystals – no structural phase changes were found to accompany changes in magnetization and polarization. The influence of epitaxy on the LG formalism can then be accounted for by adding a rhombic perturbation of K$_{pert}$ to the uniaxial magnetic anisotropy K$_u$. A phase transition from the cycloidal to the homogeneous antiferromagnetic spin states will occur at a critical value of the perturbation $K_{pert}^c$, when the energy of the cycloidal state is equal to that of the homogeneous one (see Section I.2). This will occur when the anisotropy constant fulfills the critical perturbation condition of

$$K_{pert}^c > \frac{(\gamma P_s)^2}{2A} + K_u - \frac{M_s^2}{2\chi_\perp} \approx 2 \times 10^6 \ \frac{erg}{cm^3}. \qquad (30)$$

Analysis has shown that the in-plane epitaxial constraint in (111)$_c$ BiFeO$_3$ thin films is sufficient to break the cycloidal spin order [53].

Notably higher magnetizations of 10emu/g have been reported in (001)$_c$ epitaxial layers [13]. We know for sure that the latent magnetization of the AFM state can be released: it is much lower than reported in [13], but still exceeds by more than one order the value of that induced by magnetic field of several Tesla in bulk single crystals. The possibility exists that ME films with much higher magnetizations might be engineered by crystal-chemical design [59]: however, this remains controversial [60, 61]. Interestingly, it has recently been shown [62] that La substitution (5 at %) can increase the magnetization by a factor of two in BiFeO$_3$; and up to 3.5emu/g for (BiFeO$_3$)$_{1-x}$–(PbTiO$_3$)$_x$ films [63]. The substitution of Bi with Tb results in ferromagnetic magnetization of 600 G independent of the film thickness [64]. The influence of rare earth substitution on magnetic properties of BiFeO$_3$ is discussed in more detail in Section II.2.

The enhanced multiferroic properties of BiFeO$_3$ appear not only as large electric polarization and magnetization but also in enhanced magnetoelectric effect of 3V/(cm·Oe) [13]. There were several models to explain this phenomenon: an enhancement due to existence of $P_iL_jM_k$ interaction that renormalizes the magnetoelectric effect [48], and epitaxial constraint [53,55]: but it is still remains an open question.

In addition, as BiFeO$_3$ is a rhombohedral perovskite, it crystallizes in the same structure as several known half-metallic ferromagnets: such as La$_{2/3}$Sr$_{1/3}$MnO$_3$, La$_{2/3}$Ca$_{1/3}$MnO$_3$, and Sr$_2$FeMoO$_6$. This makes it possible to combine BiFeO$_3$ with these other perovskites in multifunctional epitaxial heterostructures. Such heterostructures could be used in spintronics as magnetic and ferroelectric tunnel junctions, controllable by electric and magnetic fields [18]. Recently, epitaxial bilayers integrating a BiFeO$_3$ layer on a La$_{2/3}$Sr$_{1/3}$MnO$_3$ bottom electrode and a SrTiO$_3$ substrate were successfully grown [65]. Structural, electrical and magnetic properties of said heterorostructures were studied; and it was shown that the magnetic properties of La$_{2/3}$Sr$_{1/3}$MnO$_3$ are preserved and that the BiFeO$_3$ layers are insulating and ferroelectric down to thicknesses of 5 nm. Clearly, BiFeO$_3$ ultra-thin layers have the potential to fulfill some important criteria for use as ferroelectric tunnel barriers. Moreover, the antiferromagnetic domain switching induced by electric field has been demonstrated in BiFeO$_3$ thin films very recently [66] that provide the new possibility for tunnel switching devices[2].

## II.2 Effect of rare earth substituents.

Compounds with the formula RFeO$_3$ (i.e., rare-earth orthoferrites) also have a perovskite structure; though, orthorhombically distorted. The introduction of rare-earth substituents can thus change the anisotropy constant, so that the presence of spatially modulated structures would be energetically unfavorable. The first experiments of this type were done in the late 1980s - early 1990s on crystals of Bi$_{1-x}$R$_x$FeO$_3$ (R =*La, Gd, Dy*) for 0.4<*x*<0.5 [67-69], which revealed the presence of a linear ME effect up to liquid nitrogen temperatures, as shown in Figure 10a. The high magnetic field measurements showed that the presence of lanthanum additives x~0.1 reduces the transition field from the spatially modulated state to homogenous [70,71]. Direct evidence of cycloid suppression in La-substituted bismuth ferrite ceramics was provided by nuclear magnetic resonance [72]. The NMR spectra for Bi$_{1-x}$La$_x$FeO$_3$ for x=0.1, 0.2, 0.3 are shown in Figure 10b. It is clearly seen that the doubled NMR line (corresponding to that of the cycloid) transforms into a single broad one (corresponding to that of the homogenous antiferromagnetic state) with increasing La content.

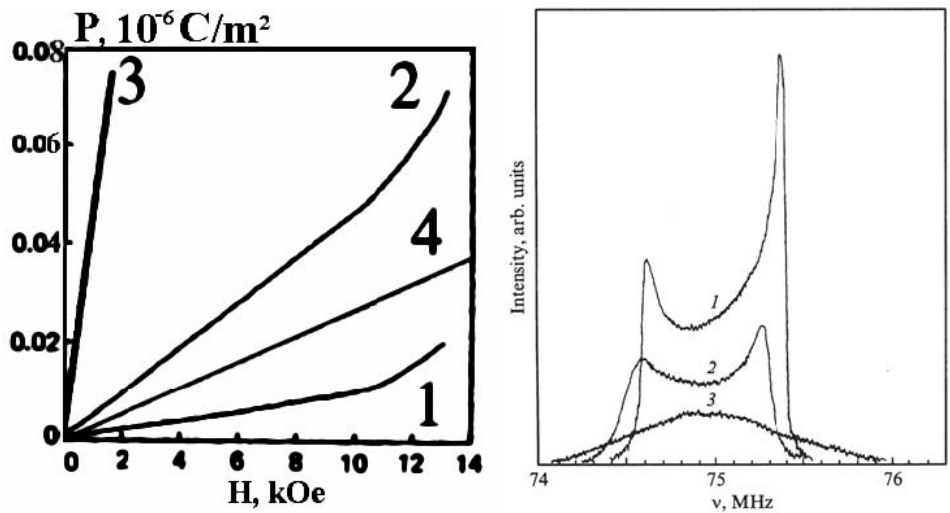

*Figure 10. (a) Magnetoelectric effect in the orthorhombic crystals Bi$_{1-x}$R$_x$FeO$_3$ (plane 001) H*⊥*(001) (1) Bi$_{0.45}$La$_{0.55}$FeO$_3$ T=4.2K (2) Bi$_{0.55}$Gd$_{0.45}$FeO$_3$ T=4.2K (3) Bi$_{0.55}$Dy$_{0.55}$FeO$_3$ T=4.2K (4) Bi$_{0.45}$Dy$_{0.55}$FeO$_3$ T=77K; [69] (b) Spin cycloid suppression in Bi$_{1-x}$La$_x$FeO$_3$ with increasing of La content: (1) x=0 (2) x=0.1 (3) x=0.2 [72].*

In search of BiFeO$_3$-based materials with enhanced multiferroic properties [73-76], attention has also be drawn to solid solutions of BiFeO$_3$–*x*PbTiO$_3$ ceramics substituted with La on the Bi sites. For Bi$_{0.8}$La$_{0.2}$FeO$_3$-43%PbTiO$_3$ [75], an anomalously high polarization (see

---

[2] This text is dated 2006. Nowdays the electric field control of magnetization is implemented in exchanged coupled structures of Co$_{0.9}$Fe$_{0.1}$ ferromagnetic layer on the BiFeO$_3$ antiferromagnetic substrate, see *Y.H. Chu et al* Nature Materials, v. 7, p. 478 (this comment is made in 2008).

Fig.11a) and the appearance of a small spontaneous magnetization (see Fig.11b) have been reported in the vicinity of a morphotropic phase boundary (MPB) between rhombohedral and tetragonal ferroelectric phases, about which numerous monoclinic and an orthorhombic bridging phases have also been reported in related ferroelectric perovskites [77-81].

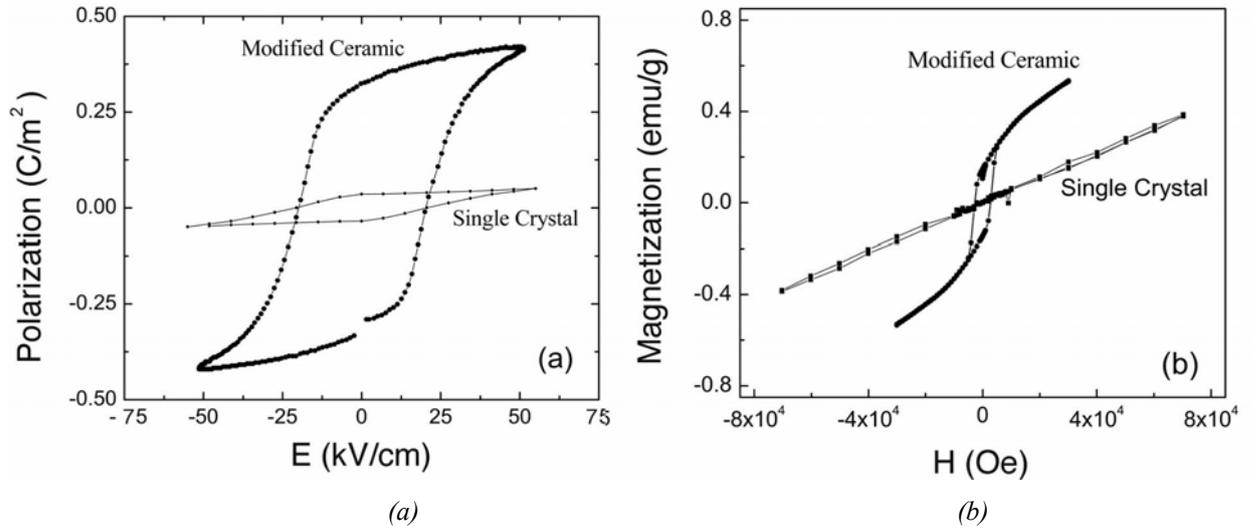

*Figure 11. Polarization and magnetization behaviors for $Bi_{0.8}La_{0.2}FeO_3$-43%$PbTiO_3$ ceramics: (a) P-E response; and (b) M-H response. Data for a $[111]_c$ oriented $BiFeO_3$ single crystal is also shown for comparison [75].*

It is interesting to note that MPB solid solutions of $BiFeO_3$–$xPbTiO_3$ were initially reported in 1962, at the time of the earliest multiferroic studies [77]. However, enhanced multiferroic properties were not reported until recently. Reminding us that persistence is important in the search for high performance single-phase ME materials!

## Part III. Lesson(s) learned, and new horizons: *'Zhong young zhi dao'*, 中庸之道

The multiferroic properties of $BiFeO_3$ result from 'compromises'. The weak antiferromagnetism of the homogeneous spin state only arises due to magnetoelectric interactions with an internal effective electric field created by the polarization. This weak magnetization is not expressed in the zero field condition in bulk crystals, rather prefers to '*sleep quietly*' within a long-period incommensurately modulated spin structure. It is worth noting that, in general, incommensuration itself results from '*compromises*' between competing interactions [30]. This homogeneous spin state can be '*awakened*' by ordering fields that perturb the symmetry of the magnetocrystalline anisotropy, upon: (i) application of high magnetic fields, in bulk crystals; (ii) imposing epitaxial constraints, in thin layers; and (iii) local orthorhombic distortions (or random field-like interactions) induced by rare-earth substituents. Thus, the lesson that we should have learned is that of '*Zhong young zhi dao*': the ancient Chinese 'doctrine of the mean or middle grounds'.

We can cite laminate ME composites [82] and self-assembling nano-composites as examples of new horizons offered by understanding this '*middle ground*'. Composites assembled (at length scales ranging from nm to mm) from two phases of entirely different symmetries and electronic configurations: allowing polarization and magnetization to each have '*their best space*'. Thus, systems which reveal the coexistence of large polarizations and large magnetizations can be fabricated. In the case of laminated composites, magnetoelectric susceptibilities as high as $\alpha_{me}$=5x$10^{-7}$s/m [or 50V/cm-Oe] have been reported near the electromechanical resonance frequency, which is nearly five orders of magnitude higher than that of the value for classical magnetoelectric $Cr_2O_3$. In the mentioned (nano- or macro-) composite systems, the magnetoelectric properties are not restricted by the transformation characteristics of Table II, but rather are product tensor properties created by the elastic '*co-* *operation*' between the magnetostriction of the magnetic phase and the electrostriction of the polar one.

We conceptually illustrate this approach in Figure 12, following work of Zheng et. al. [82], which shows a self-assembled nanocomposite consisting of $CoFe_2O_4$ nano-rods quasi-periodically dispersed in a $BaTiO_3$ matrix. This epitaxial film was grown on $SrRuO_3/SrTiO_3$ electrode/substrate by pulsed laser deposition from an initially two-phase ceramic target. The structures have different lattice parameters, and thus this on-growth morphology is naturally created to minimize the elastic energy. The left-hand figure shows the polarization hysteresis loop for the $BaTiO_3$ matrix, and the right-hand one shows the magnetization hysteresis loop for the $CoFe_2O_4$ nano-rods[3]. Clearly, large polarizations and magnetizations co-exist on an intimate length scale in this artificially architectured film. The bottom part of the figure clearly shows that the spontaneous magnetization experiences a change on going through the ferroelectric Curie temperature: demonstrating exchange between the polarization and magnetic subsystems.

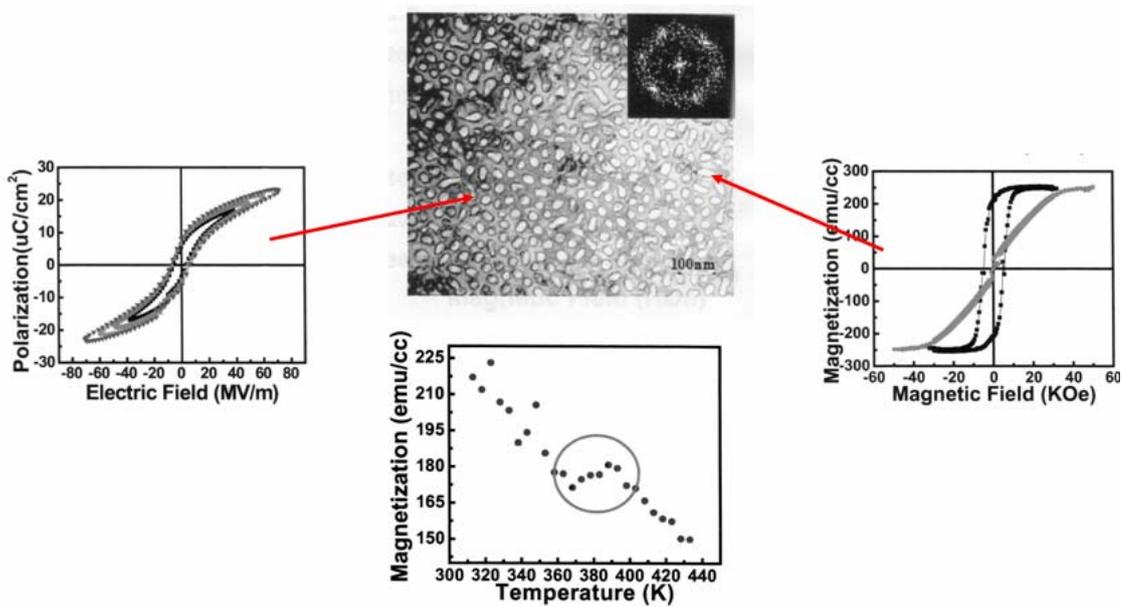

*Figure 12. Illustration of the morphology and properties of self-assembled eptiaxial $CoFe_2O_4$ nano-rods embedded in a $BaTiO_3$ matrix, both deposited epitaxially on top of a $SrRuO_3/SrTiO_3$ electrode/substrate. Top figure: bright field image of the nano-composite morphology; left-hand figure: P-E response of $BaTiO_3$ matrix; right-hand figure: M-H response of $CoFe_2O_4$ nano-rods; and bottom figure: magnetization as a function of temperature, illustrating a change near the ferroelectric Curie temperature[82].*

Clearly, composites offer a 'middle grounds' approach to room temperature ME materials. The system is not single phase; although the morphological regularity of Fig.12 has some limited characteristics of a lattice arrangement of phases. New horizons lay in this nano-scale, and not just thin-layered ones that naturally assemble under epitaxial constraint; but, rather truly architecturally-engineered systems are conceivable, if and when they can be made. Using said approach, we could envision bulk materials in which ferromagnetic and ferroelectric phases are intimately mixed on fine scales, in a manner allowing for continuity of both flux lines, creating an effective medium that appears to be single phase to electromagnetic radiation.

Finally, we illustrate several simple architecturally-engineered nano-structured $BiFeO_3$ layers. First, by varying the oxygen pressure during deposition, the dominant phase in a $BiFeO_3$

---

[3] It is worth noticing that self assembled $BiFeO_3/CoFe_2O_4$ columnar nanostructures demonstrate not only ferroelectric and ferromagnetic hysteresis but also electric field-induced magnetization switching, see *F. Zavaliche et al*, Nano Letters, **5**, 1793 (2005)

film has been shown to continuously change from BiFeO$_3$ to a mixture of α-Fe$_2$O$_3$ and γ-Fe$_2$O$_3$. This represents a new type of multiferroic nano-composite: where α-Fe$_2$O$_3$ and γ-Fe$_2$O$_3$ are magnetostrictive and BiFeO$_3$ piezoelectric [83]. Enhancement of the ME effect in the case of small ME particles embedded in a dielectric matrix with large value of the dielectric constant has recently been considered [84]. Second, the synthesis and characterization of ordered multiferroic BiFeO$_3$ nanotube arrays has been reported [85]. Nanotubes with diameters of about 250 nm, wall thickness of 20 nm, and lengths of about 6 μm were fabricated by a sol-gel method utilizing nano-channel alumina templates, as shown in Figure 13. Investigations have shown that BiFeO$_3$ maintains its ferroelectric properties in this morphological form: however, investigations have yet to be performed to exploit this architecture, and characterize its structure-property relations.

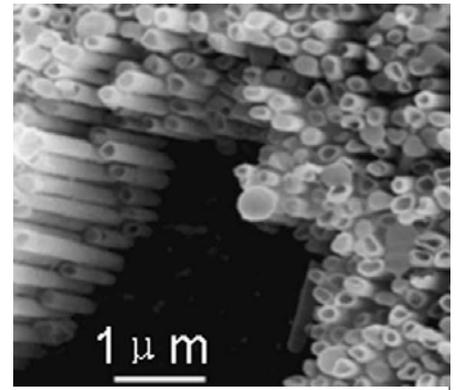

*Fig. 13 SEM image of BiFeO$_3$ nanotube arrays. Reprinted with kind permission of J.Y. Dai et al [85].*

**References**

[1] G. A. Smolenskii and I. E. Chupis, Sov. Phys. Usp. **25**, 475(1982).
[2] J. B. Neaton, C. Ederer, U. V. Waghmare, N. A. Spaldin, and K. M. Rabe, Phys. Rev. B **71**, 014113 (2005)
[3] E. A. Turov, Usp. Fiz. Nauk **164**, 325 (1994) [Phys. Usp.**37**, 303 (1994)].
[4] A.M. Kadomtseva, A.K. Zvezdin, Yu. F. Popov, A.P. Pyatakov and G.P. Vorob'ev, JETP Lett. **79**, 571 (2004).
[5] V.M. Dubovik, Phys. Rep. **187,** 145 (1990)
[6] V. L. Ginzburg, A. A. Gorbatsevich, Yu. V. Kopaev, and B. A. Volkov, Solid State Commun. **50**, 339 (1984).
[7] Yu. F. Popov, A. M. Kadomtseva, S. S. Krotov et al., Fiz. Nizk. Temp. 27, 649 (2001) [Low Temp. Phys. **27**, 478(2001)]
[8] H. Schmid, Ferroelectrics **252**, 243/41 (2001).
[9] I. Sosnowska, T. Peterlin-Neumaier, and E. Steichele, J. Phys. C **15**, 4835 (1982)
[10] M.Agaltsov, V.S.Gorelik, A.K.Zvezdin, V.A.Murashov and D.N.Rakov, Lebedev Physics Institute Letters **5**, 37 (1989).
[11] J. R. Teague, R. Gerson, and W. J. James, Solid State Commun. **8**, 1073 (1970).
[12] P. Fischer, M. Polomska, J. Phys. C: Solid State **13,** 1931(1980)
[13] J Wang., H. Zheng, V.Nagarajan et al., Science **299**, 1719 (2003)
[14] A. K. Zvezdin and A. P. Pyatakov, Physics-Uspekhi. **47**, 8 (2004).
[15] M. Fiebig, J. Phys. D: Appl. Phys. **38,** R123 (2005).
[16] W. Eerenstein, N. D. Mathur, J. F. Scott, Nature, **442**, 759 (2006)
[17] C. Binek, A. Hochstrat, X. Chen, P. Borisov and W. Kleemann, J. Appl. Phys. **97**, 10C514 (2005); Ch. Binek and B. Doudin, J. Phys.: Condens. Matter **17**, L39 (2005).
[18] M. Gajek, M. Bibes, A. Barthelemy, K. Bouzehouane, S. Fusil, M. Varela, J. Fontcuberta, and A. Fert, Phys. Rev. B **72**, 020406R (2005)
[19]A.S. Logginov, A.P. Pyatakov and A.K. Zvezdin, Proc. SPIE **5955**, 565 (2005).

[20] G. Smolenskii, V. Isupov, A. Agranovskaya, and N. Krainik, Sov. Phys. Solid State **2**, 2651 (1961).
[21] S. V. Kiselev, R. P. Ozerov, and G. S. Zhdanov, Dokl. Akad. Nauk SSSR **145**, 1255 (1962) [Sov. Phys. Dokl. **7**, 742 (1962)].
[22] F. Kubel and H. Schmid, Acta Crystallogr. **B 46**, 698 (1990).
[23] I. E. Dzyaloshinskii, Zh. Éksp. Teor. Fiz. **32**, 1547 (1957) [Sov. Phys. JETP **5**, 1259 (1957)]
[24] F. Moriya, Phys. Rev. **120**, 91 (1960).
[25] C. Tabares-Munoz, J.–P. Rivera, A. Bezinges, A. Monnier, and H. Schmid, Jap. J. Appl. Phys. **24**, suppl. 24-2, 1051 (1985).
[26] A.V. Zalessky, A. A. Frolov, T.A. Khimich, A.A. Bush, V.S. Pokatilov and A.K. Zvezdin, Europhys. Lett. **50**, 547 (2000).
[27] R Przeniosło, A Palewicz, M Regulski, I Sosnowska, R M Ibberson, and K S Knight, J. Phys.: Condens. Matter **18**, 2069 (2006).
[28] Yu. F. Popov, A. K. Zvezdin, G. P. Vorob'ev *et al.*, Pis'ma Zh. Eksp. Teor. Fiz. **57**, 65 (1993) [JETP Lett. **57**, 69 (1993)].
[29] Yu. F. Popov, A. M. Kadomtseva, G. P. Vorob'ev, and A. K. Zvezdin, Ferroelectrics **162**, 135 (1994).
[30] I. Sosnowska and A. K. Zvezdin, J. Magn. Magn. Mater. **140–144**, 167 (1995).
[31] V. G. Bar'yakhtar, V. A. L'vov, and D. A. Yablonskii, JETP Lett. **37**, 673(1983).
[32] A. A. Khalfina and M. A. Shamtsutdinov, Ferroelectrics, **279**, 19 (2002).
[33] A. S. Logginov, G. Meshkov, A. V. Nikolaev, and A. Pyatakov, JETP Lett. **86**, 115 (2007)
[34] M. Mostovoy, Phys. Rev. Lett. **96,** 067601 (2006).
[35] A. Sparavigna, A. Strigazzi and A. Zvezdin, Phys. Rev. B **50** 2953 (1994).
[36] A.K. Zvezdin, Bulletin of the Lebedev Physics Institute.**4**, 5 (2002) Allerton Press Inc/New York)
[37] L. Néel and R. Pauthenet, C. R. Acad. Sci., Paris **234**, 2172 (1952).
[38] A. S. Borovik-Romanov and M. P. Orlova, Zh. Éksp. Teor. Fiz. **31**, 579 (1956) [Sov. Phys. JETP **4**, 531 (1956)].
[39] I. E. Dzyaloshinskii, Zh. Éksp. Teor. Fiz. **37**, 881 (1959) [Sov. Phys. JETP **10**, 623 (1960)].
[40] D. N. Astrov, Zh. Éksp. Teor. Fiz. **38**, 984 (1960) [Sov.Phys. JETP **11**, 708 (1960)].
[41] V. J. Folen, G. T. Rado, and E. W. Stalder, Phys. Rev. Lett. **6**, 607 (1961).
[42] I. E. Dzyaloshinskiii, Zh. Éksp. Teor. Fiz. **33**, 807 (1957) [Sov. Phys. JETP **6**, 621 (1958)].
[43] A. S. Borovik-Romanov, Zh. Éksp. Teor. Fiz. **38**, 1088 (1960) [Sov. Phys. JETP **11**, 786 (1960)].
[44] A. S. Borovik-Romanov and B. E. Javelov, Proc.3rd Reg. Conf. Magn. (Prague, 1963), p. 81
[45] N. F. Kharchenko, A. V. Bibik, and V. V. Eremenko, Pis'ma Zh. Éksp. Teor. Fiz. **42**, 447 (1985) [JETP Lett.**42**, 553 (1985)].
[46] V. S. Ostrovskii and V. M. Loktev, Pis'ma Zh. Éksp. Teor. Fiz. **26**, 139 (1977) [JETP Lett. **26**, 130 (1977)].
[47] S. Leykuras, H. Legal, D. Minella, *et al.*, Physica B (Amsterdam) **89**, 43 (1977).
[48] A.K. Zvezdin, Bulletin of the Lebedev Physics Institute.**4**, 3 (2004).
[49] Tehranchi M.M., Kubrakov N.F., Zvezdin A.K., Ferroelectrics, **204**, 181 (1997)

[50] A. G. Zhdanov, A. K. Zvezdin, A. P. Pyatakov, T. B. Kosykh, and D. Viehland, Phys.Solid State **48,** 88 (2006).
[51] B. Ruette, S. Zvyagin, A.P. Pyatakov, A. Bush, J.F. Li, V.I. Belotelov, A.K. Zvezdin, and D. Viehland, Phys. Rev. B 69, 064114 (2004).
[52] Jiefang Li, Junling Wang, M. Wuttig, R. Ramesh, Naigang Wang, B. Ruette, A. P. Pyatakov, A. K. Zvezdin, and D. Viehland, Appl. Phys. Lett. **84**, 5261 (2005).
[53] F. Bai, J. Wang, M. Wuttig, J.F. Li, N. Wang, A. Pyatakov, A.K. Zvezdin, L.E. Cross, D. Viehland, Appl. Phys. Lett. **86**, 032511 (2005).
[54] Guangyong Xu, H. Hiraka, G. Shirane, Jiefang Li, Junling Wang and D. Viehland, Appl. Phys.Lett. **86**, 182905 (2005).
[55] Q. Jiang, and J. H. Qiu, J. Appl. Phys. **99**, 103901 (2006).
[56] Kwi Young Yun, Minoru Noda, and Masanori Okuyama, Appl. Phys. Lett. **83**, 3981 (2003)
[57] Xiaoding Qi, Joonghoe Dho, Rumen Tomov, Mark G. Blamire, and Judith L. MacManus-Driscoll, Appl. Phys. Lett. **86**, 062903 (2005).
[58] Can Wang et al., J. Appl. Phys. **99**, 054104 (2006).
[59] J. Wang et al., Science **307,** 1203b (2005).
[60] W. Eerenstein et al Science **307,** 1203a (2005).
[61] H. Béa, M. Bibes, A. Barthélémy, K. Bouzehouane, E. Jacquet, A. Khodan, and J.-P. Contour S. Fusil, F. Wyczisk, A. Forget, D. Lebeugle, D. Colson, and M. Viret, Appl. Phys. Lett. **87**, 072508 (2005).
[62] Dongeun Lee, Min G. Kim, Sangwoo Ryu, and Hyun M. Jang, Sang G. Lee, Appl. Phys. Lett. **86**, 222903 (2005).
[63] Liu Hongri, Liu Zuli, Liu Qing and Yao Kailun, J. Phys. D: Appl. Phys. **39**, 1022 (2006)
[64] V. R. Palkar, K. Ganesh Kumara, and S. K. Malik, Appl. Phys. Lett. **84**, 15 (2004).
[65] H. Béa, M. Bibes, M. Sirena, G. Herranz, K. Bouzehouane, and E. Jacquet, S. Fusil, P. Paruch and M. Dawber, J.-P. Contour and A. Barthélémy, Appl. Phys. Lett.. **88**, 062502 (2006).
[66] T. Zhao, A. Scholl, F. Zavaliche, et al., Nature Materials 5 823 (2006).
[67] V. A. Murashev, D. N. Rakov, I. S. Dubenko, A.K.Zvezdin, V.M.Ionov, Kristallografiya **35**, 912 (1990) [Sov. Phys. Crystallogr. **35**, 538 (1990)].
[68] V.A.Murashov, D.N.Rakov, N.A.Ekonomov, A.K.Zvezdin and D.N.Dubenko, Solid State Physics (Leningrad) **32**, 2156 (1990).
[69] Z. V. Gabbasova, M. D. Kuz'min, A. K. Zvezdin, I.S. Dubenko, V.A. Murashev, D.N. Rakov and I.B. Krynetsky, Phys. Lett. A **158**, 491 (1991).
[70] G. P. Vorob'ev, A. K. Zvezdin, A. M. Kadomtseva *et al.*, Fiz. Tverd. Tela (St. Petersburg) **37**, 2428 (1995) [Phys.Solid State **37**, 1329 (1995)].
[71] A. M. Kadomtseva, Yu. F. Popov, G. P. Vorob'ev, and A. K. Zvezdin, Physica B **211**, 327 (1995).
[72] V. Zalesskii, A. A. Frolov, T. A. Khimich, and A. A. Bush, Phys. Solid State **45**, 141 (2003) (FTT . 45, 135)
[73] J.S. Kim, C. Cheon, Y.N. Choi and P.W. Jang, J. Appl. Phys. **93**, 9263 (2003).
[74] V. R. Palkar, Darshan, C. Kundaliya, S.K. Malik and S. Bhattacharya, Phys. Rev. B **69,** 212102 (2004).
[75] N. Wang, J. Cheng, A. Pyatakov, A.K. Zvezdin, J.F. Li, L.E. Cross, D. Viehland, Phys. Rev. B 72, 1 (2005).

[76] G. L. Yuan and Siu Wing Or, Appl. Phys. Lett. **88**, 062905 (2006).

[77] S.A.Fedulov,Yu.N.Venevtsev and G.S.Zhdanov, Kristallografiya **7**, 77 (1962) (in Russian).

[78] J.Cheng, W.Zhu, N.Li,and, L.E.Cross, Mater. Lett. **57**, 2090 (2003).

[79] Jinrong Cheng, R. Eitel, and L.E. Cross, J. Am. Ceram. Soc. **86**, 2111 (2003).

[80] Jinrong Cheng and L.E.Cross, J.Appl.Phys. **94**, 5188 (2003).

[81] R.E. Eitel, C.A. Randall, T.R. Shrout and S.-E.Park, Jpn. J.Appl. Phys. **41**, 2099 (2002).

[82] H. Zheng, J. Wang. S. Lofland, Z. Ma, T. Zhao, S. Shinde, S. Ogale, F. Bai, D. Viehland, Y. Jia, D. Schlom, M. Wuttig, A. Roytburd, and R. Ramesh, Science **303**, 661 (2004).

[83] M. Murakami, S. Fujino, S.-H. Lim, L. G. Salamanca-Riba, M. Wuttig, I. Takeuchi, Bindhu Varughese, H. Sugaya, T. Hasegawa and S. E. Lofland, Appl. Phys. Lett. **88**, 112505 (2006).

[84] A.K. Zvezdin, Bulletin of the Lebedev Physics Institute.**4**, 28 (2004).

[85] X. Y. Zhang, C. W. Lai, X. Zhao, D. Y. Wang, and J. Y. Dai, Appl. Phys. Lett. **87**, 143102 (2005).